\pdfoutput=1
\documentclass[
 reprint,
 amsmath,amssymb,
 aps,
]{revtex4-1}

\usepackage{graphicx}
\usepackage{dcolumn}
\usepackage{bm}
\usepackage{float}

\begin{document}

\title{Observation of Optical Backflow}

\author{Yaniv Eliezer$^\ddagger$}
\author{Thomas Zacharias$^\ddagger$}
\author{Alon Bahabad}
\email{alonb@eng.tau.ac.il}
\affiliation{Department of Physical Electronics, School of Electrical
	Engineering, Fleischman Faculty of Engineering,\\ and Center of
	Light-Matter Interaction, \\ Tel-Aviv University, Tel-Aviv 69978, Israel}

\date{\today}

\begin{abstract}
	Backflow is a counter-intuitive phenomenon in which a forward propagating quantum particle propagates locally backwards. The actual counter-propagation property associated with this delicate interference phenomenon has not been observed to date in any field of physics. Here, we report the observation of transverse optical backflow where a beam of light propagating to a specific transverse direction is measured locally to propagate in the opposite direction. This observation is relevant to any physical system supporting coherent waves and might lead to unique applications.   
\end{abstract}

\pacs{Valid PACS appear here}

\maketitle

\section {\label{sec:level1} Introduction}
	Backflow (also known as retro-propagation) is a surprising yet relatively unknown phenomenon. It was first pointed out in 1969 \cite{allcock1969time1,allcock1969time2,allcock1969time3} by G. R. Allcock in the context of the time-of-arrival problem in quantum theory. Allcock found that a local quantum probability current may become negative even for positive momenta quantum states, and thus cannot be a valid measure for the time-of-arrival. Further advances with regards to the time-of-arrival problem were made by Muga \cite{muga1999arrival, muga2000arrival,damborenea2002measurement}.
	The phenomenon was studied in detail in 1994 by Bracken and Melloy \cite{bracken1994probability} who found a limit on the total amount of backflow. This led them to introduce a new dimensionless quantum number whose value has been reproduced more accurately in subsequent years \cite{eveson2005quantum, penz2005new}. Recently there has been a renewed interest in backflow with various studies reintroducing and exploring various aspects of the phenomenon \cite{yearsley2013introduction, albarelli2016quantum}. 
	
	A recent development in the field has included an exploration of the relation between the phenomena of superoscillations, weak measurements, and backflow \cite{berry2010quantum}. Superoscillation is the phenomenon in which a band-limited signal locally oscillates faster than its fastest Fourier component. The phenomenon appeared first in the microwave community \cite{schelkunoff1943mathematical,slepian1961prolate,di1952super}
	and later revived in quantum theory by Aharonov et al. in 1988 \cite{aharonov1988result} while establishing the theory of quantum weak measurements. Additional theoretical works by Berry and Popescu developed the mathematical theory of the phenomenon and further proposed to use superoscillations to realize far-field sub-wavelength optical focusing without evanescent waves \cite{anandan1994quantum, berry2006evolution}.
	This suggestion was verified experimentally shortly afterward \cite{huang2007focusing}. Various experimental works involving superoscillations, which appeared in the last two decades, includes super-resolution microscopy \cite{rogers2013optical, huang2007optical,wong2013optical}, optical beam shaping \cite{Greenfield2013,eliezer2016super,zacharias2017axial}, nano-focusing of light \cite{David2015}, particle trapping \cite{singh2017particle}, electron beam shaping \cite{remez2017superoscillating} and nonlinear optical frequency conversion \cite{remez2015super}. Optical superoscillations were also demonstrated experimentally in the time domain \cite{eliezer2017breaking,eliezer2018experimental}. A complementary phenomenon to superoscillation, termed suboscillation, where a lower bound limited signal oscillates  locally slower than its lowest Fourier component, was discovered recently \cite{eliezer2017super}.
	
	The connection between superoscillation and backflow was first noted by Berry \cite{berry2010quantum}, where he analyzed the evolution of backflow regions in the interference of quantum wavepackets. He demonstrated that these regions are always dependent on the overall momenta distribution of the wavepackets and showed that the superposition of many waves creates wider and stronger backflow regions when the Fourier components are highly correlated. Moreover he also found this phenomenon to be extremely vulnerable, since the evolution in space-time causes the destruction of the delicate phase relations which are critical for backflow. In 2013, Palmero et al. \cite{palmero2013detecting} proposed an experiment to detect quantum backflow by applying a Bragg pulse to a Bose-Einstein Condensate. A recent work investigated the effect of reflection and transmission processes on backflow \cite{PhysRevA.96.012112}. Importantly, the first experimental observation of the local momentum  associated with backflow near optical superoscillatory foci was reported \cite{yuan2018plasmonics}. Still, to the best of our knowledge, no experimental observation of any actual backward movement associated with backflow in any wave system has been reported to date.
	 
	\begin{figure*}[t] 
		\begin{centering}
			\makebox[\textwidth][c]{
				\includegraphics[width=1.0\linewidth,trim={3.3cm 2.2cm 6cm 1.5cm},clip]{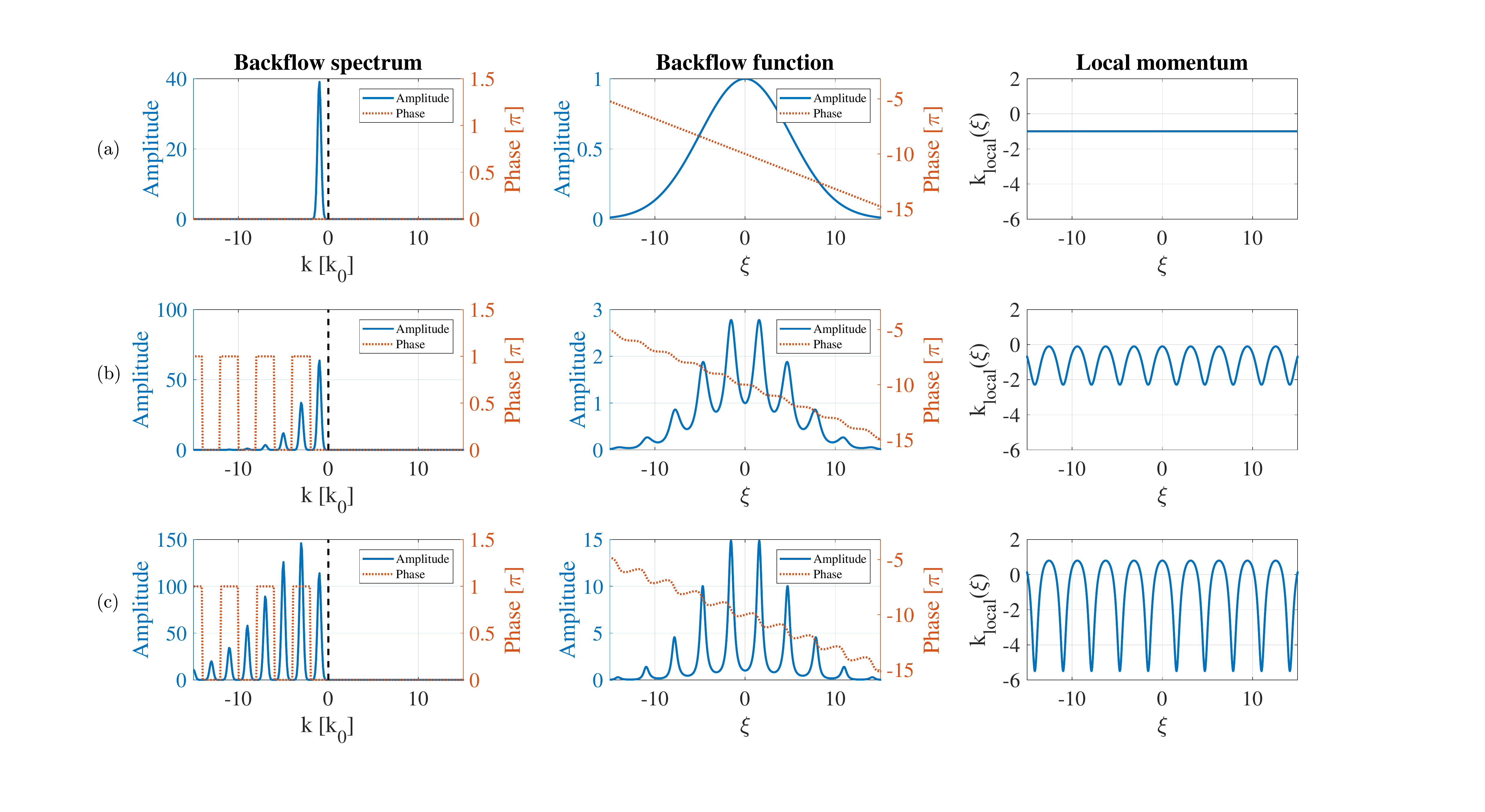}
			}
			\par\end{centering}
		\caption{ \textbf{The finite backflow function $\mathbf{f_{FBF}(\xi)}$}. \textbf{(Left)} The backflow function spatial spectrum. $k_0$ is the fundamental spatial frequency. The dashed black line represents the centre of the $k$ axis, related to zero transverse momentum. \textbf{(Centre)} The backflow function in real space. \textbf{(Right)} The local spatial frequency of $f_{FBF}(\xi)$. The rows correspond to the cases of \textbf{(a)} $a = 1.0$, \textbf{(b)} $a = 0.7$ , and \textbf{(c)} $a = 0.4$.
		}
		\label{fig:theory}
	\end{figure*}
	
	In this work we experimentally observe counter-propagation due to optical backflow. We construct a light beam, based on a spectrally shifted suboscillatory function \cite{eliezer2017super}, made of the superposition of modes having negative transverse momentum relative to a chosen axis of propagation.  While the expectation value of the transverse momentum of the beam is negative (i.e. the beam travels at a negative angle relative to the chosen axis of propagation), in certain locations the local value of the transverse momentum is positive. Isolating these regions with a slit causes the local transverse momentum to be ``projected'' onto the expectation value of the beam's transverse momentum, realizing a beam propagating at an overall positive angle relative to the propagation axis.

\section{\label{sec:level2} Theory}

	Consider the following backflow function built using a spectrally-shifted suboscillatory function $f_{Sub}(\xi)$ \cite{eliezer2017super} in the coordinate $\xi$:
	\begin{equation}
		f_{BF}(\xi) = f_{Sub}(\xi)\exp(iL\xi) = \dfrac{\exp(iL\xi)}{\left[\cos (k_0\xi) + ia \sin (k_0\xi)\right]^N},
		\label{eq:bfeqn}
	\end{equation}
	where $a \in \{0<\mathbb{R} \le 1\}$, $N \in \{\mathbb{N}>0\}$, $k_0$ is the fundamental spatial frequency of the suboscillatory function and $\exp(iL\xi)$ acts as a spectral shifting function. The Fourier transform of $f_{BF}(\xi)$ is given with:
	\begin{equation}
		F_{BF}(k) = 2\pi \sum_{m = -\infty}^{+\infty}C_m(a)\cdot\delta(k - L - m k_0),
		\label{eq:bfeqnFourierTrans}
	\end{equation}
			
	where $k$ denotes the spatial frequency, $C_m(a)$ are the Fourier coefficients and $L$ is the spectral shift parameter.

	By applying the local frequency operator on Eq. (\ref{eq:bfeqn}), we get: 
	\begin{equation}
		k_{local}(\xi) = Im{\frac{\partial \ln[f_{BF}(\xi)]}{\partial \xi}} = L - \dfrac{N a k_0}{\cos^2(k_0 \xi) + a^2 \sin^2(k_0 \xi)}.
		\label{eq:localfrequency}
	\end{equation}
	
	According to Eq. (\ref{eq:localfrequency}) it is obvious that $k_{local}(\xi) \in \left[L-\frac{N k_0}{a},L-N k_0 a\right]$. The global spectrum of $f_{Sub}(\xi)$ is a negatively-valued single sided spectrum with the highest harmonic at $-N k_0$ \cite{eliezer2017super} (which sets it as the slowest frequency component). For the function $f_{BF}(\xi)$, the $\exp({iL\xi})$ term shifts the highest available spectral frequency to $M=-N k_0+L$. 
	This implies that for a proper selection of the parameters $L$, $N$, $k_0$, and $a$  it is possible to obtain a single sided spectrum composed of only negative components $M < 0$ while having a local positive frequency $(L-N k_0 a)>0$, which is the hallmark of the backflow phenomenon.
	Consider for example the Fourier coefficients of Eq. (\ref{eq:bfeqn}) equipped with $N = 3$, which are calculated by complex integration to be: 
	\begin{equation}
		{C_m}(a) = \left\{ {\begin{array}{*{20}{c}}
			{\left( {{m^2} - 1} \right)\frac{{{{\left( {a + 1} \right)}^{\frac{m}{2} - \frac{3}{2}}}}}{{{{\left( {a - 1} \right)}^{\frac{m}{2} + \frac{3}{2}}}}}},&{m \in \left\{ {{\rm{odd}} < 0} \right\}},\\
			0,&{\textrm{otherwise.}}
			\end{array}} \right .
		\label{eq:bfeqnFourierCoeff}
	\end{equation}	
	
	Together with $k_0 = 1$ and $L=2$ the spectrum in Eq. (\ref{eq:bfeqnFourierTrans}) is completely single sided, having negative only components, with the highest harmonic at $M = -1$. The local frequency, however, as defined by Eq. (\ref{eq:localfrequency}), is positive in the regions close to $\xi = 2\pi n$ ($n \in \mathbb{N}$) for $a < 2/3$.
	
	The backflow function in Eq. (\ref{eq:bfeqn}) is periodic and extends from $-\infty$ to $\infty$ and so cannot be used in an experiment. We therefore derive a bounded signal by convolving the Fourier transform in Eq. (\ref{eq:bfeqnFourierTrans}) equipped with the coefficients of Eq. (\ref{eq:bfeqnFourierCoeff}) with a narrow Gaussian spectral distribution. The resulting spectrum of this finite backflow function is given with:
	\begin{eqnarray}
	F_{FBF}\left( k \right) = \sum\limits_{m =  - P}^{-N} {{C_m^{}(a)}\exp\left( { - \frac{{{{\left[ {k - L - {m k_0}} \right]}^2}}}{{2{\sigma _0}^2}}} \right)},
	\label{gioexpsum}
	\end{eqnarray} 	
	where $C_m(a)$ are the coefficients defined in (\ref{eq:bfeqnFourierCoeff}), $L$ is the spectral shift parameter, $k_0$ is the fundamental spatial frequency and ${\sigma_0}^2$ stands for the spectral variance of each harmonic. Since $C_m(a)$ decay rapidly, we synthesize our function using a finite number of negative harmonics, where $(-P)$ and $(-N)$ represent the indices corresponding to the lowest and highest harmonics. A sufficiently narrow spectral variance is chosen such that theoretically only a negligible fraction ($3.82 \times 10^{-12}$) of the energy of the highest mode $M$ is positive. Experimentally, see below, we realize our beam using discrete sampling of the theoretical smooth function for which there is no energy at all associated with the Gaussian envelopes at positive transverse frequency values. The resulting spectrum is shown in Fig. \ref{fig:theory} (left) for several values of the parameter $a$. The inverse Fourier transform of this spectrum, which corresponds to the finite backflow function $f_{FBF}(\xi)$, is shown in Fig. \ref{fig:theory} (centre). Fig. \ref{fig:theory} (right) shows the local momentum calculated by Eq. (\ref{eq:localfrequency}) for each case. It can be seen that for the case of $a = 0.4$, while the entire spectrum resides in the negative side (Fig. 1 (c)(left)), positive local momentum regions are clearly evident (Fig. 1 (c)(right)).

\section{\label{sec:level3} Experiment}

\begin{figure}[h] 
		\begin{centering}
			\includegraphics[width=1.0\linewidth,trim={4.7cm 3cm 5.6cm 7.5cm},clip]{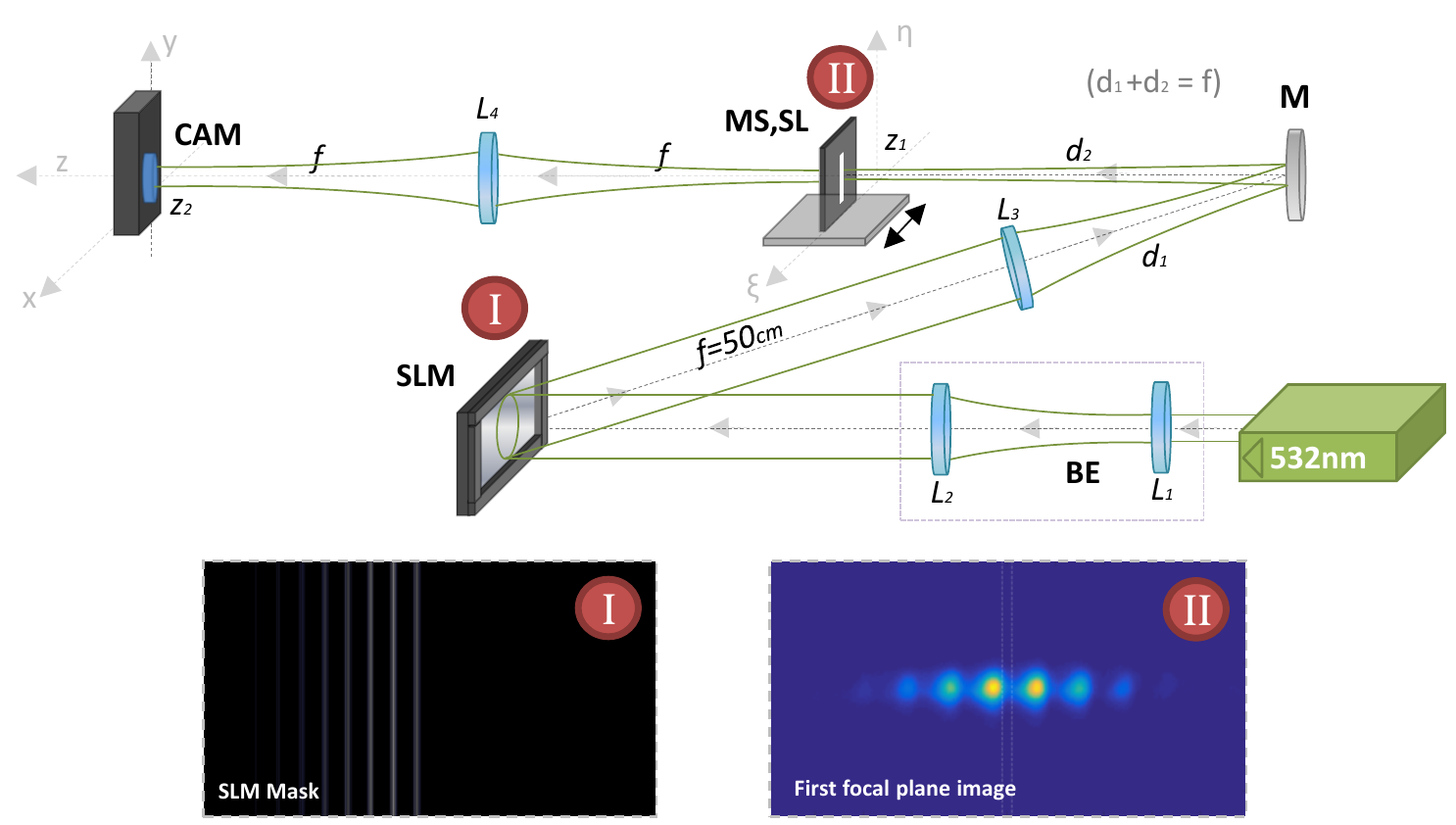}
			\par
		\end{centering}
		\caption{ \textbf{Experimental setup.} BE = beam expander. SLM = spatial light modulator. MS = moving stage. SL = slit. M = mirror. $L_1, L_2, L_3, L_4$ are lenses.  $d_1+d_2$ equals to lens $L_3$ focal length. $Z_1$ marks the location of the 1st focal plane and $Z_2$ marks the location of the 2nd focal plane. \textbf{(Inset)} \textbf{(I)} A realization of one of the phase-only masks used in the experiment. \textbf{(II)} A realization of the first-diffraction-order beam intensity at the 1st focal plane.
		}
		\label{fig:setup}
	\end{figure}

	Our experimental setup (see Fig. \ref{fig:setup}) consists of a $532$ nm continuous-wave laser (Quantum Ventus $532$ Solo Laser) and a reflective phase-only SLM (Holoeye Pluto SLM) on which we mount our optical masks. An integrated part of the optical mask is a grating whose 1st diffraction order constitutes the propagation direction associated with the transverse frequency $k=0$. The laser beam is expanded and collimated before the SLM, reflected off it and Fourier transformed using a $50$ cm focal lens into the 1st focal plane. The generated first-diffraction-order beam was imaged at the 1st focal plane before being passed through a $100 \mu m$ wide optical slit (Thorlabs S100RH) placed at various positions along the horizontal axis in the 1st focal plane. The position of the slit was controlled using a $25 mm$-long stepper motor linear stage (Newport MFA-PPD) which was set to move in steps of $5 \mu m$. The spatially filtered beam in the 1st focal plane was then Fourier transformed again by a $50$ cm focal lens into the 2nd focal plane where the intensity was imaged. All images were taken using a CMOS camera (Ophir Spiricon SP620U).  
	 
\begin{figure*}[t] 
			\begin{minipage}[b]{1.0\linewidth}			
				\begin{centering}
					\makebox[\textwidth][c]{
						\includegraphics[width=0.9\linewidth,trim={3.8cm 1.0cm 4.0cm 0.5cm},clip]{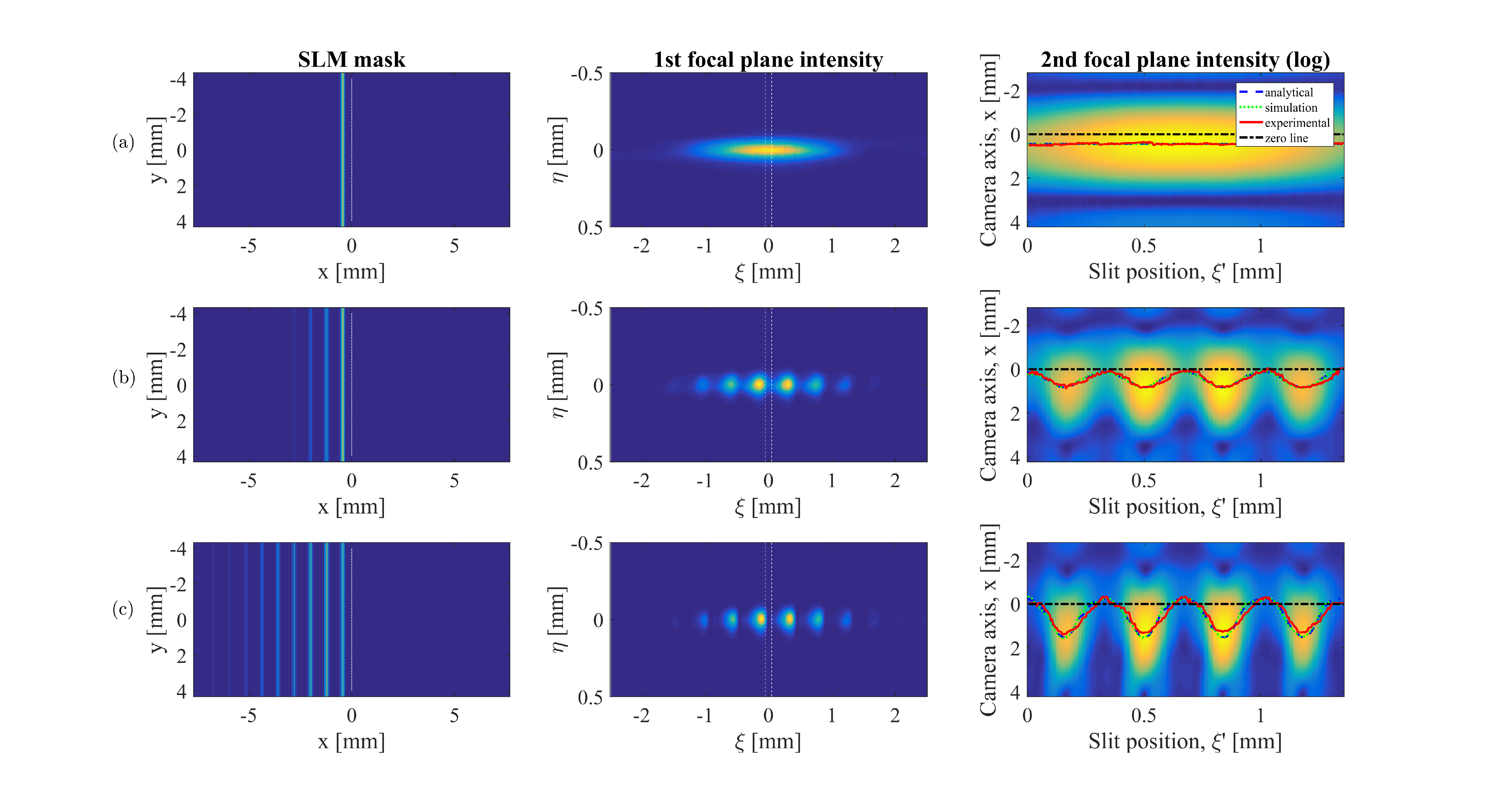}
					}
					\par\end{centering}
				\caption{ \textbf{Experimental measurements.} \textbf{(Left)} The generated SLM phase-only masks. Each line creates a propagating mode with a well-defined negative transverse momentum. The dotted line represents the centre of the $x$ axis, related to zero transverse momentum.  \textbf{(Center)} Measured intensity distribution in the 1st focal plane, which is essentially the backflow beam. The two dashed white lines represent the width of the slit used to scan the beam. \textbf{(Right)} The measured beam image at the 2nd focal plane, averaged over the $y$ coordinate for each slit position. The dashed-dotted black line denotes the centre of the propagation axis. Continuous red line: measured expectation value of the beam position (equal to momentum in the 1st focal plane). Dashed blue line: analytically calculated expectation value for a theoretical infinite periodic backflow beam after it is slit-filtered. Dotted green line: expectation value derived from Fourier transforming the SLM image and then Fourier-transforming again the slit-filtered image. \textbf{(a)} $a = 1$,  \textbf{(b)} $a = 0.7$, \textbf{(c)} $a = 0.4$.
				}
				\label{fig:experiment}
			\end{minipage}
		\end{figure*}

	To demonstrate optical backflow we first calculated the spectrum of the finite backflow function for different values of the tuning parameter $a = \{0.4, 0.5, 0.6, 0.7, 0.8, 0.9, 1\}$, with $P=21$, $N=3$, and $L=0.733mm^{-1}$, according to Eq. (\ref{gioexpsum}). The fundamental frequency and each harmonic spectral variance were selected to be $k_0 = 0.391mm^{-1}$ and $\sigma_0 = 0.0611mm^{-1}$ correspondingly. The resulting complex-valued spectral functions were encoded into a set of phase-only masks on the SLM using a known phase-encoding technique \cite{bolduc2013exact}.	To isolate the desired beam from the unmodulated residual field, the beam was sent to the 1st diffraction order using masks containing phase-only blazed gratings having a period of $\Lambda = 64\mu m$. The encoded masks corresponding to the values of $a = 0.4$, $a = 0.7$, $a = 1$ appear on Fig. \ref{fig:experiment} (left). To calibrate the location of the propagation axis at the 2nd focal plane, a single slit mask was placed at the centre of the SLM screen. The expectation value of the measured image in the 2nd focal plane determines the location of the propagation axis in that plane for all subsequent measurements. 	
	
	Each of the phase masks generated a finite backflow field pattern at the 1st focal plane. The pattern's intensity was imaged and can be seen in Fig. \ref{fig:experiment}(centre). Then a $100\mu m $ moving slit was placed at the 1st focal plane. The slit scanned the backflow patterns along the $\xi$ axis. The slit-filtered beam then propagated through another $2f$ system before being imaged at the 2nd focal plane, for each slit position. By calculating the expectation value of the beam's position at the 2nd focal plane we find the degree of the beam's deflection relative to the propagation axis, for each position of the slit. Notice that we start in the SLM plane with a momentum representation of the backflow function and end back at a plane representing momentum. Thus the beam deflection represents expectation values of the transverse momentum.  
	
	The right column of Fig. \ref{fig:experiment} describes the measured deflection for different positions of the slit. Note that the application of an optical Fourier transform twice using two consecutive $2f$ systems results with a coordinate inversion. The curves in Fig. \ref{fig:experiment} represent the expectation of the beam's distribution in the 2nd focal plane. They were calculated from the captured beam images (red continuous line), compared to an analytical expression based on the finite backflow beam after it was slit-filtered (dashed blue line), and to a numerical calculation based on two consecutive optical Fourier transforms of the SLM patterns (dotted green line). All the results show very high level of agreement. A comparison of the expectation curve in the cases of $a = 1$ (a), $a = 0.7$ (b), and $a = 0.4$ (c) indicates that, as expected, the degree of deflection due to backflow increases as the value of $a$ decreases (and the beam does not contain backflow for $a=1$). We also note that the $a=0.4$ case (c) achieves the maximal backflow value in the sense that the beam's deflection completely crosses the propagation axis for certain positions of the slit, thus materializing  local positive momentum. It is also clear that larger backflow entails lower intensity in the beam which is a defining characteristic of super and sub-oscillating functions  \cite{aharonov2010time,aharonov2017mathematics,tamir2013introduction,1742-6596-70-1-012016,berry2011pointer}. 
	
	\begin{figure*}[ht] 
		\begin{minipage}[b]{1.0\linewidth}					
			\begin{centering}
				\includegraphics[width=0.75\linewidth,trim={3.5cm 0.0cm 3.0cm 0.0cm},clip]{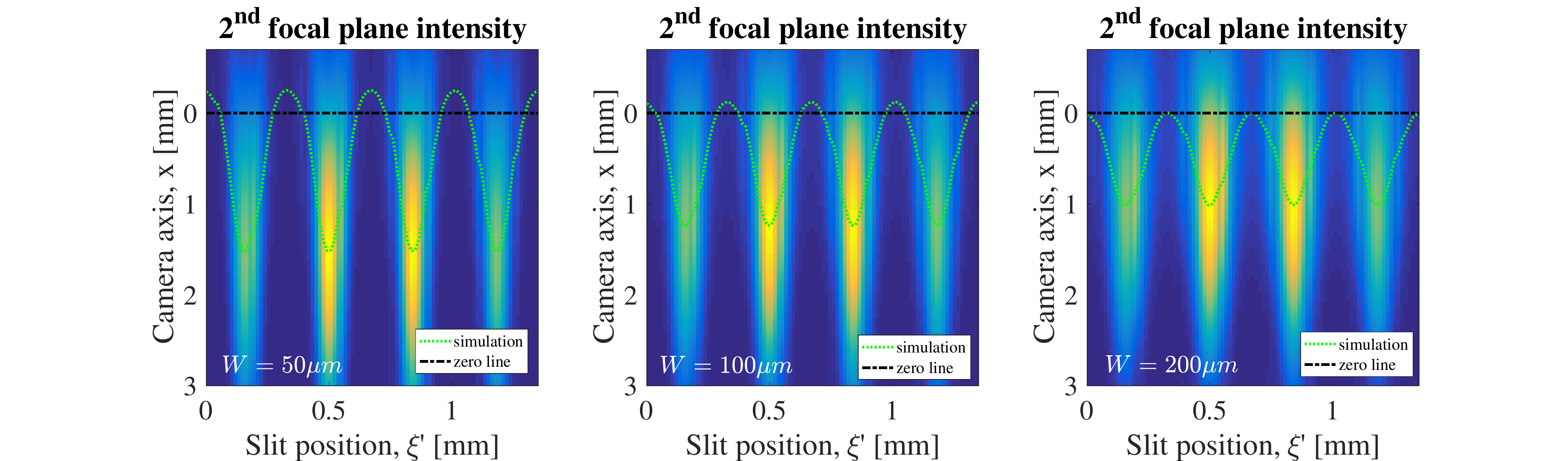}
				\par
			\end{centering}
			\caption{ \textbf{Intensity distribution for different slit widths}. 
				Simulation of the 2nd focal plane intensity distribution expectation with respect to the $y$ axis as a function of the slit's position, for the case of $a = 0.4$, and for different values of the slit's width: \textbf{(Left)} $W = 50\mu m$, \textbf{(Centre)}  $100\mu m$, and \textbf{(Right)} $200\mu m$. 
				The dotted green curve represents the beam's position expectation in both axis.
			}
			\label{fig:IntenVWidth}
		\end{minipage}		
	\end{figure*}	
	
	In a sense the filtering by the slit applied on the backflow beam, realizes a nonlinear ``projection''  operation of a local property (local transverse momentum, which is not an eigenvalue of the momentum operator $\hat {k} \rightarrow -i\partial_x$) to a global property (eigenvalue of the momentum operator)  thus allowing to observe the backflow as an actual deflection of the beam. 
	This is related to the formalism of weak measurements \cite{berry2010quantum}:  the local momentum is considered as the result of a weak measurement, giving rise to observable $A_{weak}$ through the operation of the momentum operator $\hat{k}$ on a preselected state which is, in our case, the backflow beam $\psi$ (whose representation in momentum space is $\langle k | \psi \rangle=F_{FBF}(k)$ and in coordinate space $\langle \xi | \psi \rangle=f_{FBF}(\xi)$) and after post selecting with a coordinate state $|\xi\rangle$ (describing the position of the slit) \cite{berry2010quantum}:  
	\begin{equation}
		k_{local}(\xi) =  Im{\frac{\partial \ln[\psi(\xi)]}{\partial \xi}} = Re\frac{\langle \xi|\hat{k}|\psi\rangle}{\langle \xi|\psi \rangle} =  A_{weak}.
		\label{eq: localMomentumWeak}
	\end{equation}

	 To examine the deflection's sensitivity to the slit's width, $W$, we have used a beam propagation simulation to calculate the beam's deflection as a function of the slit's position and width. Figure \ref{fig:IntenVWidth} shows the expectation value of the beam's position with respect to the $y$ axis, in the 2nd focal plane, for different values of the slit's width: $W = 50\mu m$ (a), $W = 100 \mu m$ (b), and $W = 200\mu m$ (c), all cases with the backflow parameter $a=0.4$. The green dotted curves represent the expectation value of the beams' position along $x$. It is evident from the curves that the degree of backflow decreases as the slit width increases. Of course, reducing the slit width increases the variance in the position of the beam (its width), but the change in the position of the centre of mass of the beam (its expectation value) is due to the local momentum at the slit position.
	
\section{\label{sec:level4} Conclusions}

	We experimentally constructed, measured and observed beam deflections associated with backflowing beams. We first designed a backflow beam based  on the mathematical form of suboscillatory functions which were discovered recently. This allowed controlling  the degree of backflow in the beam. Slit-filtering the generated beams allowed, counterintuitively, their deflection towards a direction opposite to that associated with the momentum states comprising the original beam. This effect is the result of a delicate interference phenomenon which, until now, hindered the observation of movement or deflection associated with backflow in any wave system (from quantum particles to optical waves to acoustic waves etc.).      
	The consequences of this proof of principle experiment might be far reaching for possible applications. For example, it can be very interesting to use backflow for stand-off spectroscopy \cite{coffey2013advances,gottfried2009laser} where optical sensing of remote areas is based on information propagating backwards towards the source. It might also be relevant to nonlinear and ultrafast optics especially if this phenomenon is manifested in the time domain, generating local negative frequencies. Finally as already noted, being a universal wave phenomenon this effect is relevant to any wave system and it might find unique uses in various fields of physics.  

\section{Author contribution}
A.B. conceived the idea and supervised the study. Y.E. performed the analytical analysis and numerical simulations. Y.E. and T.Z. performed the experiments. The manuscript was written by all co-authors. Y.E. and T.Z. contributed equally to this work.

\section{Data availability}
The data and findings of this study are available from the corresponding authors upon a reasonable request.

\section{Competing interests}
The authors declare no competing interests.

\bibliographystyle{ieeetr}
\bibliography{bibliography}

\end{document}